\documentclass[12pt]{article}
\usepackage{amsfonts,amssymb}

\advance\hoffset by -1.1truecm
\advance\voffset by -2.0truecm

\makeatletter
\@ifundefined{newsymbol}
{
\renewcommand{\AA}{\mbox{\rm\hspace{.22em}%
\vrule width .02em height 0.6ex \hspace{-.24em}A}}
\newcommand{\BB}{\mbox{\rm I\hspace{-.22em}B}}
\newcommand{\C}{\mbox{C \hspace{-1.16em} \raisebox{-0.018em}{\sf l}}\;}
\newcommand{\HH}{\mbox{I\hspace{-0.22em}H}}

}{
\renewcommand{\AA}{{\Bbb A}} 
\newcommand{\BB}{{\Bbb B}} 
\newcommand{\C}{{\Bbb C}} 
\newcommand{\HH}{{\Bbb H}} 
}
\makeatother

\newcommand{\opname}[1]{\mathop{\rm #1}\nolimits} 

\newcommand{\Cl}{\opname{C\ell}} 
\newcommand{\End}{\opname{End}} 
\newcommand{\Hom}{\opname{Hom}} 
\newcommand{\tr}{\opname{tr}} 
\newcommand{\Wres}{\opname{Wres}} 

\newcommand{\A}{{\cal A}} 
\renewcommand{\a}{\alpha} 
\renewcommand{\b}{\beta} 
\newcommand{\Ca}{{\cal C}} 
\renewcommand{\d}{{\rm d}} 
\newcommand{\del}{\partial} 
\newcommand{\Dslash}{D\mkern-11.5mu/\,} 
\newcommand{\eqn}[1]{{\rm(\ref{#1})}} 
\newcommand{\Ga}{\Gamma} 
\newcommand{\ga}{\gamma} 
\renewcommand{\H}{{\cal H}} 
\renewcommand{\L}{{\cal L}} 
\newcommand{\la}{\lambda} 
\newcommand{\nc}{{\mathrm{nc}}} 
\newcommand{\ox}{\otimes} 
\newcommand{\sepword}[1]{\qquad\mbox{#1}\quad} 
\newcommand{\sol}{{\sf so}} 
\newcommand{\stroke}{\mathbin{\vert}} 
\newcommand{\thalf}{{\textstyle\frac{1}{2}}} 
\newcommand{\tihalf}{{\textstyle\frac{i}{2}}} 
\newcommand{\tquarter}{{\textstyle\frac{1}{4}}} 
\newcommand{\w}{\wedge} 
\newcommand{\x}{\times} 
\renewcommand{\:}{\colon} 

\def\<#1,#2>{\langle#1\stroke#2\rangle} 

\newbox\ncintdbox \newbox\ncinttbox
\setbox0=\hbox{$-$} \setbox2=\hbox{$\displaystyle\int$}
\setbox\ncintdbox=\hbox{\rlap{\hbox to \wd2{\hskip-.125em\box2\relax
\hfil}}\box0\kern.1em}
\setbox0=\hbox{$\vcenter{\hrule width 4pt}$}
\setbox2=\hbox{$\textstyle\int$}
\setbox\ncinttbox=\hbox{\rlap{\hbox to \wd2{\hskip-.175em\box2\relax
\hfil}}\box0\kern.1em}
\newcommand{\ncint}{\mathop{\mathchoice{\copy\ncintdbox}%
{\copy\ncinttbox}{\copy\ncinttbox}{\copy\ncinttbox}}\nolimits}

\newcommand{\CMP}[3]{Commun. Math. Phys. {\bf #1} (19#2) {#3}}
\newcommand{\JGP}[3]{J. Geom. Phys. {\bf #1} (19#2) {#3}}
\newcommand{\JMP}[3]{J. Math. Phys. {\bf #1} (19#2) {#3}}
\newcommand{\PNAS}[3]{Proc. Nat. Acad. Sci. USA {\bf #1} (19#2) {#3}}
\newcommand{\PRL}[3]{Phys. Rev. Lett. {\bf #1} (19#2) {#3}}
\newcommand{\RMP}[3]{Rev. Mod. Phys. {\bf #1} (19#2) {#3}}
\newcommand{\journal}[4]{#1 {\bf #2} (19#3) {#4}}

\title{A nonperturbative form of the spectral action
principle in noncommutative geometry}

\author{H. Figueroa,$^1$ J. M. Gracia-Bond\'{\i}a,$^{1,2}$
F. Lizzi$^{3,4}$ and J. C. V\'arilly$^{1,5}$\thanks{%
E-mail: \texttt{figo@cariari.ucr.ac.cr}, \texttt{lizzi@na.infn.it},
\texttt{varilly@cpt.univ-mrs.fr}}\\[12pt]
$^1$Departamento de Matem\'atica, Universidad de Costa Rica,\\
2060 San Pedro, Costa Rica\\[3pt]
$^2$Departamento de F\'{\i}sica Te\'orica, Universidad de Zaragoza,\\
50009 Zaragoza, Spain\\[3pt]
$^3$Dipartimento di Scienze Fisiche, Universit\`a di Napoli
{\it Federico II},\\ and INFN, Sezione di Napoli, Italy\\[3pt]
$^4$Theoretical Physics, 1 Keble Road, Oxford OX1 3NP, UK\\[3pt]
$^5$Centre de Physique Th\'eorique, CNRS--Luminy, Case 907,\\
13288 Marseille, France}

\date{}

\begin{document}

\begin{titlepage}

\maketitle
\thispagestyle{empty}

\begin{abstract}
Using the formalism of superconnections, we show the existence of a
bosonic action functional for the standard $K$-cycle in noncommutative
geometry, giving rise, through the spectral action principle, only to
the Einstein gravity and Standard Model Yang--Mills--Higgs terms. It
provides an effective nonminimal coupling in the bosonic sector of the
Lagrangian.
\end{abstract}

\vspace{1cm}

\begin{flushright}
\begin{tabular}{r@{\qquad}r}
CPT--97/P.3453 & DFTUZ/97/03\\
DSF--8/97 & OUTP--97--05P\\
UCR--FM--10--97 & hep-th/9701179
\end{tabular}
\end{flushright}

\end{titlepage}

\subsection*{Introduction}

The Connes--Lott models~\cite{Book,RealNCG,Cordelia} of noncommutative
geometry (NCG) have so far yielded action functionals both for
elementary particles (tying together the gauge bosons and the Higgs
sector) and for gravity \cite{ConnesLH,KastlerEH,KalauW}. The
challenge of unifying the Yang--Mills and gravitational actions was
taken up in \cite{ConnesGrav} and subsequently Chamseddine and Connes
put forward a model \cite{ChamC,ChamCbis} doing so by means of a
so-called universal action functional. Their approach is based on the
hypothesis that fundamental interactions are coded in the invariants
of a suitably generalized Dirac operator, involving spacetime and
internal variables. Needless to say, this bold introduction of
spectral geometry in physics has important consequences, even for
classical relativity~\cite{LandiR}.

In the Chamseddine--Connes (CC) as in the Connes--Lott Ans\"atze,
particle species are taken as given. Setting the fermionic action is
equivalent to fixing a ``real $K$-cycle'' $(D,\ga,J)$ comprising the
generalized Dirac operator, grading and conjugation for the theory. To
this one adds a bosonic action $B[D]$ depending solely on~$D$ (though
implicitly also on~$\ga$ and~$J$). The choice of $B$ thus becomes the
critical issue for the physical interpretation. Papers
\cite{ChamC,ChamCbis} start from the $K$-cycle
currently~\cite{RealNCG,Cordelia} associated to the Standard Model (or
standard $K$-cycle) and concentrate on aspects that depend only weakly
on that choice; thus the adjective ``universal''.

The CC approach has two important merits, namely the possibility of a
genuine unification of particle theories and gravity and the
introduction of a renormalization process to control the mix of
physical scales involved. Nonetheless, there are some difficulties
that remain to be addressed. First of all, the particular approach
taken by Chamseddine and Connes raises mathematical questions about
the information content of the asymptotic developments
used~\cite{Odysseus}. Secondly, their action has a number of extra
terms one could do without. The leading term is a huge cosmological
constant that has to be ``renormalized away'' with fine tuning. The
gravity part of the third term contribution, comprising a Weyl gravity
term and a term coupling gravity with the Higgs field, is conformally
invariant. It is unclear at present if the latter is more an asset or
a liability in black hole dynamics and in cosmology~\cite{EYepes}.

Also, the renormalization scheme proposed in \cite{ChamC,ChamCbis}
exhibits some surprising traits. The CC Lagrangian, as it stands,
being a higher-derivative theory without the $R^2$ term, is neither
renormalizable {\it strictu sensu}, nor unitary within the usual
perturbation theory~\cite{Enrique,BuchbinderOS}. The first objection
is not considered serious in the modern effective field approach to
quantum field theory~\cite{Donoghue}. Nonunitarity is a quantum
analogue of ill-posedness of the classical Cauchy problem (for the
hyperbolic version of the Lagrangian). This second objection is
dismissed on the grounds that we expect the product geometry to be
replaced by a truly noncommutative geometry at some energy scale lower
than the Planck mass. Chamseddine and Connes chose a cutoff scale of
the order $10^{15}$~GeV, running in conflict with the value of
Newton's constant. It is fair to say that we really do not know the
energy scale at which the NCG relations can claim validity. The fact
that the most natural coefficients for the boson fields they obtain
yield $SU(5)$-type relations for the chromodynamical and
flavourdynamical coupling constants is perhaps not enough of an
indication, as the theory still lacks a physical unifying mechanism at
the $10^{15}$~GeV or other scale.

Models based on the ``universal'' functional concept are also
aesthetically unappealing to some. One can hardly help being
mesmerized by the beauty of the results of \cite{KastlerEH,KalauW}, in
which a particular regularized functional, the Wodzicki residue of the
inverse squared (ordinary) Dirac operator, gives directly the
Einstein--Hilbert functional for gravity. The idea of then keeping the
Wodzicki functional and further modifying the Dirac operator, in such
a way that all the action terms of the Standard Model plus gravity
---and only them--- are obtained, was proposed by Ackermann
\cite{Ackermann} and spelled out recently by
Tolksdorf~\cite{Tolksdorf}.

The Ackermann-Tolksdorf (AT) formalism ---that falls outside NCG---
has its own drawbacks, however: their manipulation of the Dirac
operator physically amounts to a nonminimal coupling of the fermions
and the gauge fields, that contradicts the present tenets of quantum
field theories. In the fermion sector, this form of nonminimal
coupling would give rise to a coupling between two fermions and two
bosons, which has never been seen. Moreover, the fermion doubling
demonstrated in \cite{DoubleTrouble} is compounded.

Underlying Ackermann and Tolksdorf's attempt, there is perhaps the
impression that a pure, combined Einstein--SM Lagrangian cannot be
obtained from NCG. Such an impression seems widespread ---and indeed
the original Chamseddine and Connes' papers did not clarify the point
either. But it is not in correspondence with the facts: in this paper
we show that one can, at least in principle, obtain the pure
Einstein--SM Lagrangian at the tree level from the same standard
$K$-cycle used by Chamseddine and Connes. Whether the extra terms
present in the CC perturbative development are a necessity or not is
to be decided by quantum field theoretical considerations and/or
experiment.

\subsection*{Action functionals in NCG}

A NCG model is determined by an algebra $\A$ having a representation
on a Hilbert space $\H$, on which there also act a grading operator
$\ga$, a conjugation $J$ and a selfadjoint operator $D$, odd with
respect to~$\ga$ and commuting with~$J$, with suitable properties
vis-a-vis the algebra; in particular one requires that the operators
$[D,a]$ commute with $JbJ^{-1}$, for $a,b$ in~$\A$. This five-term
package~\cite{RealNCG,Cordelia} is called a ``spectral triple'' or a
``real $K$-cycle''.

As stated in \cite{ConnesGrav}, a commutative $K$-cycle is just the
spectral version of a Riemannian spin manifold (a compact spacetime,
able to uphold fermions, with Euclidean signature). Let $M$ be such a
manifold, with dimension $n$; we take $\A = C^\infty(M)$,
$\H = L^2(S)$, the space of square-integrable spinors over~$M$,
$\ga = \ga_5$, $J$ is charge conjugation of the spinors and
$D = \Dslash = \ga^a(\del_a + \omega_a)$, where $\omega$ is the spin
connection 1-form, is the ordinary Dirac operator on~$M$. The metric
tensor on~$M$ (and then its functionals) is completely determined
by~the $K$-cycle.

At the other extreme, $\A = \A_F$ could be finite-dimensional but
noncommutative, $\H_F$ also finite-dimensional and graded, $D_F$ an
odd matrix; this $K$-cycle describes a (noncommutative) internal
space. In the applications to the Standard Model the entries of $D_F$
are Yukawa--Kobayashi--Maskawa parameters: they are seen as part and
parcel of the geometry.

All $K$-cycles employed in NCG till now are ``mildly noncommutative''
product $K$-cycles, where $\A = C^\infty(M) \ox \A_F$,
$\H = L^2(S) \ox \H_F$ and the ``free'' Dirac operator is given by
$D_f = \ga^a\del_a \ox 1 + 1 \ox D_F$. We call the second piece a
Dirac--Yukawa operator. To turn the NCG machinery, one needs to
introduce the noncommutative gauge potential $A_\nc$, a general
selfadjoint element of the form $\sum a[D_f, b]$ corresponding to the
``fluctuations'' of the internal degrees of freedom. To this one adds
the spin connection (only known way of incorporating the fermions into
the external geometry). For the standard $K$-cycle,
$\A_F = \HH \oplus \C \oplus M_3(\C)$, and in that way one reproduces
the fermionic part of the Standard Model Lagrangian, with an important
amount of fermion doubling, however, that needs to be projected out to
get the physical fermion sector.

According to the spectral action principle, the bosonic action depends
on the whole $K$-cycle; we shall write $B[D]$ for short. One
postulates the Lagrangian density
$$
\L = \<\psi, PDP\psi> + B[D],
$$
where $P$ projects on the subspace of the physical Weyl fermions. We
shall concentrate on $B[D]$. This bosonic part in the original model
and its subsequent modifications was fabricated following a
``differential'' path as follows: given the noncommutative gauge
potential $A_\nc$, one constructed its curvature
$F_\nc = [D, A_\nc] + A_\nc^2$ (a far from straightforward task, due
to the ambiguity of the NCG differential structure), and the action
was taken to be proportional to $\ncint F^2 \,ds^4$ (see further on
for the definition of the noncommutative integral $\ncint\,$). In the
case of the standard $K$-cycle, this indeed defines the usual
Yang--Mills action and the action for the Higgs field, including the
usually ad hoc quartic potential. Thus ``low energy'' particle
interactions were unified in a single term, the square of a geometric
object, excluding Einstein gravity.

In this paper, we follow the CC approach in exploring a fully
``integral'' path for the construction of $B[D]$.

Due to the product structure of the $K$-cycle, the fermionic states in
NCG so far always live in spaces of sections of superbundles. We
formalize this last remark. Suppose, for definiteness, that $M$ is an
even-dimensional manifold, with a spin structure; let $S$ be the
spinor bundle; write $\Cl M$ for the bundle over~$M$ whose fibre
at~$x$ is the complex Clifford algebra $\Cl(T_x^*M)$; the smooth
sections of these bundles form respectively the space of spinors
$\Ga(S)$ and the algebra $\Ca := \Ga(\Cl M)$. This algebra acts
irreducibly on $\Ga(S)$, i.e., we have $\Ca \simeq \End S$; this can
be taken as defining the spin structure~\cite{Plymen}. If we think of
$\H_F$ as the trivial bundle $\H_F \x M$, then $\H$ can be identified
to the space of sections $\Ga(S \ox \H_F)$ of the tensor product
superbundle $S \ox \H_F$. Any superbundle $E = E^+ \oplus E^-$ on
which a graded action of $\Cl M$ is defined (so $\Ca$ acts on its
space of sections) is called a Clifford module. Denote by $c$ the
action of $\Ca$ on $S$; $\a \in \Ca$ acts on $\Ga(S \ox \H_F)$ by
$$
\psi \ox \omega \mapsto c(\a)\psi \ox \omega.
$$

The passage from $S$ to $S \ox \H_F$ is a ``twisting'' of the spinor
bundle. On a spin manifold, any Clifford module $\Ga(E)$ comes from
some such twisting~\cite{BerlineGV}: by Schur's lemma, any map from
$\Ga(S)$ to $\Ga(E)$ that commutes with the Clifford action is of the
form $\psi \mapsto \psi \ox \omega$, for $\omega$ a section of the
bundle of intertwining maps $W := \Hom_{\Cl M}(S, E)$. Moreover, any
endomorphism of $\Ga(E)$ that commutes with the Clifford action is of
the form $\psi \ox \omega \mapsto \psi \ox T\omega$ for some bundle
map $T\: W \to W$; in other words,
$\End_{\Cl M} E \simeq 1 \ox \End W$. The whole matrix bundle $\End E$
is generated by the subbundle $\Cl M \simeq \End S \ox 1$, acting by
the spin representation, and by its commutator $1 \ox \End W$, so we
can write $\End E \simeq \Cl M \ox \End W$.

The analogue of the volume element in noncommutative geometry is the
operator $D^{-n} =: ds^n$. And pertinent operators are realized as
pseudodifferential operators on the spaces of sections. Extending
previous definitions by Connes~\cite{Book}, we introduce a
noncommutative integral based on the Wodzicki residue~\cite{Wodzicki}:
$$
\ncint P\,ds^{n}
:= \frac{(\thalf n - 1)!}{2(2\pi)^{n/2}} \Wres P|D|^{-n}
:= \frac{(\thalf n - 1)!}{2(2\pi)^{n/2}} \int_{S^*M}
\tr\sigma_{-n}(P|D|^{-n})(x,\xi) \,d\xi \,dx.
$$
Here $\sigma_{-n}(A)$ denotes the $(-n)$th order piece of the complete
symbol of $A$ and the numerical coefficient is good for $D = \Dslash$
and $n$ even. The Wodzicki residue is known to be the only trace on
the space of pseudodifferential operators. The noncommutative integral
$\ncint$ is a trace on spaces of operators such that their commutator
with the resolvent of $|D|^{-1}$ is of order $-2$ or less: this
includes all functions of $D$ and all pseudodifferential operators of
order 0 or less~\cite{Sirius}. The definition is justified by the fact
that
$$
\ncint f\,ds^{n} =  \int f(x) \, d^nx,
$$
for $f \in C^\infty(M)$ ---represented as a (left) multiplication
operator on $L^2(S)$. From now on we take $n = 4$.

It was natural, however, for NCG to be asked about the gravitational
interaction, which, after all, is nothing but the manifestation of the
commutative geometry of spacetime. But it turns out that to use the
operator $D_f$ or, instead, $\Dslash_f = \Dslash + 1 \ox D_F$, i.e.,
to consider the ``free'' Dirac operator as comprising the spin
connection or not, is immaterial for that purpose, as any reference to
the latter vanishes from the noncommutative gauge potential. The first
important step in the direction of connecting noncommutative geometry
with gravitational physics was carried out independently by Kastler
\cite{KastlerEH} and Kalau and Walze \cite{KalauW} who, following a
suggestion by Connes, found that the Einstein--Hilbert action is given
by
$$
\ncint \Dslash^{2}\,ds^4 \propto \Wres \Dslash^{-2}.
$$

If, instead of $\Dslash$, one uses the full, gauge covariant
$D = \Dslash_f + A_\nc + JA_\nc J^{-1}$ operator, in the hope to
describe the mix of gravity with the gauge boson interaction, one
however finds only a term proportional to the square of the Higgs
field $\phi$~\cite{KalauW}, in addition to the gravitational curvature
term. We were thus stuck in a peculiar situation: one form of the
action gave the Yang--Mills term, but not the gravitational part; the
situation was inverted for the second form of the action, which only
gives the gravitational part.

On the other hand, the Chamseddine--Connes action has terms of
different orders; the first one is essentially $\Wres \Dslash^{-4}$;
the second one is again $\Wres \Dslash^{-2}$; the third (carrying the
Weyl gravity term) and subsequent ones are not Wodzicki residues, but
generalized moments~\cite{Odysseus}. One way to see the difficulty is
that the total action contains terms such as the Riemann curvature $R$
and mass term of the Higgs potential which are quadratic in the fields
(metric-graviton, Higgs and vector bosons), while the higher order
terms such like the kinetic energy of Yang--Mills fields and the rest
of the Higgs potential are quartic or contain derivatives in the
fields.

We next demonstrate, on application of Quillen's theory of
superconnections \cite{Quillen} to the standard $K$-cycle, and
\textit{provided that the internal and external degrees of freedom can
be cleanly separated}, the existence of a functional of the $K$-cycle
containing only the Einstein--Hilbert and Yang--Mills--Higgs terms, on
the same footing. This will be the noncommutative integral of the sum
of two terms multiplied by the noncommutative geometry volume element.
Somewhat paradoxically, the kinetic-looking term (the noncommutative
integral of the square of the Dirac operator again) gives rise to the
quadratic terms, while a potential-looking term, which is the square
of the reduced superconnection, gives rise to the quartic terms.

\subsection*{Quillen's superconnections}

A key ingredient in our proposed action is the fact that the
generalized Dirac operators of product $K$-cycles arise from
superconnections that are compatible with the Clifford
action~\cite{BerlineGV}. Superconnections have been already used in
NCG in~\cite{Lee}, based on earlier work of Ne'eman and Sternberg
\cite{NeemanS}, in a slightly different context and at the Yang--Mills
level only. We now briefly describe some key features of
superconnections in reference to Dirac operators.

A superconnection on the superbundle~$E$ is \emph{any} odd linear
operator $\AA$ on the module of $E$-valued differential forms
$\A(M,E)$, graded by the sum of the grading on the scalar-valued forms
$\A(M)$ and the grading on $E$, that satisfies the Leibniz rule
\begin{equation}
[\AA, \b] = \d\b  \sepword{for} \b \in \A(M),
\label{Leibniz}
\end{equation}
where the commutator is graded. If $\nabla$ is any connection,
$\AA - \nabla$ commutes with exterior products and so is itself an
exterior product by an odd matrix-valued form:
$$
(\AA - \nabla)\,\zeta = \a \w \zeta
 \sepword{for some} \a \in \A^-(M,\End E).
$$
This yields the general recipe
$$
\AA = \a_0 + \nabla + \a_2 + \a_3 +\cdots+ \a_n
$$
where $\a_{2k} \in \A^{2k}(M,\End^- E)$ and
$\a_{2k+1} \in \A^{2k+1}(M,\End^+ E)$; we have absorbed the 1-form
component $\a_1$ in the connection. In particular, $\a_0$ is just an
\textit{odd} matrix-valued bundle map: $\a_0 \in \Ga(\End^- E)$.

The Jacobi identity shows that if $\theta$ is a matrix-valued form,
then
$$
[[\AA,\theta], \b]
= [\AA, [\theta,\b]] + (-1)^{|\theta|\,|\b|} [\d\b,\theta] = 0
$$
for any $\b$ in $\A(M)$, so $[\AA,\theta]$ is a multiplication
operator. In this way the formula
$(\AA\theta) \w \zeta := [\AA,\theta]\zeta$ serves to define the
covariant derivative $\AA\theta$ in $\A(M,\End E)$; as operators,
$\AA\theta = [\AA,\theta]$. Since $\AA$ is odd, we have
$[\AA,\AA] = 2\AA^2$, and the Jacobi identity yields
$2\,[\AA,[\AA,T]] = [[\AA,\AA],T] = [2\AA^2,T]$ for any operator~$T$.
In particular, $[\AA^2,\b] = [\AA,[\AA,\b]] = \d(\d\b) = 0$ for
any~$\b$, so $\AA^2 = F_{\AA}$ in $\A^+(M,\End E)$: this is the
\textit{curvature} of the superconnection $\AA$, and it satisfies the
Bianchi identity $\AA F_{\AA} = [\AA, F_{\AA}] = [\AA, \AA^2] = 0$.

Following \cite{BerlineGV}, we say that $\AA$ is a \textit{Clifford
superconnection} if it satisfies a second Leibniz rule, involving the
Clifford action:
\begin{equation}
[\AA, c(\b)] = c(\nabla\b)  \sepword{for each} \b \in \A(M)
\label{Leibniz-LeviC}
\end{equation}
where $\nabla$ is the Levi-Civita connection on the cotangent bundle.
On a local orthonormal basis of 1-forms $\theta_a$, one has
$\nabla_\mu \theta^a = \del_\mu \theta^a - \Ga_{\mu b}^a \theta^b$ (we
use throughout Greek indexes for coordinate bases and Latin indexes
for vierbeins). The antisymmetric matrices $\a_\mu$ with entries
$-\Ga_{\mu a}^b$ (defined on a local chart $U$) make up a Lie
algebra-valued 1-form $\a$ in $\A^1(U,\sol(T^*M))$, and
$\nabla = d + \a$ over~$U$.

The spin connection $\nabla^S$ has the property \eqn{Leibniz-LeviC}.
Locally, $\nabla^S = d + \omega = d + \dot\mu(\a)$, where
$\dot\mu\: \sol(T^*M) \to \Cl M$ is the infinitesimal spin
representation of the Lie algebra of the orthogonal group,
$\dot\mu(\a_\nu) = -\tquarter \Ga_{\nu a}^b \,\ga^a \,\ga^b$. Its
curvature is $(\nabla^S)^2 = \dot\mu(\d\a + \a \w \a) = \dot\mu(R)$,
where $R \in \A^2(M,\sol(T^*M))$ is the Riemann curvature tensor:
\begin{equation}
\mu(R) = \tquarter R_{ba\nu\sigma}\,\ga^a\,\ga^b
\, d x^\nu \w \, dx^\sigma
\sepword{with} \ga^a \equiv c(\theta^a).
\label{spin-curvature}
\end{equation}
The basic property of the spin representation~\cite{Rhea} is that
\begin{equation}
[\dot\mu(T), c(\b)] = c(T\b)
\label{spinrepn-Clif}
\end{equation}
when $\b \in \A^1(M) = \Ga(T^*M)$ and $T \in \Ga(\sol(T^*M))$. This
can be seen directly, by checking the identity
$\tquarter [\ga^a\,\ga^b, \ga^c] = \ga^{[a} \delta^{b]c}$, which
entails
$[\tquarter T_a^b \,\ga^a\,\ga^b,\,\b_c\,\ga^c] = T_a^b \b_b \,\ga^a$
if $T$ is antisymmetric. (For that, notice that the commutator
$[\ga^a\,\ga^b, \ga^c]$ is zero whenever $a = b$ or the three indices
are distinct.)

On a twisted bundle $E = S \ox W$, there is the Clifford connection
$\nabla^S \ox 1$. If $\AA$ is any Clifford superconnection, then
$\AA - \nabla^S \ox 1$ commutes with the Clifford action, and
therefore it is of the form $1 \ox \BB$ where $\BB$ is an odd operator
on $\A(M,W)$ that satisfies a Leibniz rule like \eqn{Leibniz}. In
other words, the most general Clifford superconnection is of the form
\begin{equation}
\AA = \nabla^S \ox 1 + 1 \ox \BB,
\label{Clif-sconn}
\end{equation}
where $\BB$ is any superconnection on the twisting bundle~$W$. That
is, the superconnection on a space which is the product of a
continuous Riemannian spin geometry times a (noncommutative) internal
geometry splits into the usual spin connection which acts trivially
on the internal part, plus a superconnection which acts only on the
internal part.

\subsection*{The superconnection for the standard $K$-cycle}

We can identify the algebra $\Ga(\Cl M)$ with the algebra of forms
$\A(M)$ by the isomorphism $c(\b) \mapsto c(\b)1$; the inverse map
$Q\: \A(M) \to \Ga(\Cl M)$ ---denoted $\bf c$ by~\cite{BerlineGV}, who
call it ``quantization''--- allows us to Clifford-multiply by forms.
For instance, with $\sigma^{\mu\nu}
= \thalf[c(dx^\mu), c(dx^\nu)] \equiv \thalf[\ga^\mu, \ga^\nu]$, we
have $\ga^\mu\ga^\nu\,1 = dx^\mu \w dx^\nu - g^{\mu\nu}$, so that
$$
Q(\d x^\mu \w \d x^\nu)
= g^{\mu\nu} + \ga^\mu \,\ga^\nu = \sigma^{\mu\nu}.
$$

Let $\BB = \BB_0 + \BB_{1\mu}\,dx^\mu + \BB_{2\mu\nu}\,dx^\mu \w
dx^\nu + \cdots$ be a superconnection on $W$. We can now define a
Dirac operator associated to the Clifford superconnection $\AA$ of
\eqn{Clif-sconn} by
$$
D := \Dslash \ox 1 + \BB_0 + \ga^\mu\,\BB_{1\mu}
+ \sigma^{\mu\nu}\,\BB_{2\mu\nu} + \cdots.
$$
It is clear that Dirac operators in this sense are just quantizations
of superconnections. There is a one-to-one correspondence between
Dirac operators compatible with a given Clifford action and Clifford
superconnections~\cite{BerlineGV}; for example,
$\Dslash = Q(\nabla^S)$. In particular, $D$ and $\AA$ have the same
information.

All superconnections considered in~\cite{Quillen} are of the form
$\BB_0 + \nabla$.  This goes well with Connes' formalism for product
$K$-cycles, as in the present context the superconnection pair
$(\BB_0, \BB_1)$ and noncommutative differential 1-forms are one and
the same thing ---with the degree zero term corresponding to the
Dirac--Yukawa operator. On the other hand, the AT formalism employs
superconnections with forms up to degree two.

Now, in view of equation \eqn{Clif-sconn} and the fact that
$[\nabla^S \ox 1, 1 \ox \BB] = 0$, the curvature of $\AA$ splits as
\begin{equation}
\AA^2  = \dot\mu(R) \ox 1 + 1 \ox {\BB}^2
=: \dot\mu(R) \ox 1 + 1 \ox F_{\BB}.
\label{curv-split}
\end{equation}
One can also remark~\cite{BerlineGV} that, from the Leibniz rule
\eqn{Leibniz-LeviC}:
$$
[\AA^2, c(\b)] = [\AA, [\AA,c(\b)]] = c(\nabla^2\b) = c(R\b),
$$
whereas $[\dot\mu(R), c(\b)] = c(R\b)$ from \eqn{spinrepn-Clif}. With
regard to the factorization $\End E \simeq \Cl M \ox \End W$,
$\dot\mu(R)$ acts by Clifford multiplications and we can write it as
$\dot\mu(R) \ox 1$. Thus $\AA^2 - \dot\mu(R) \ox 1$ commutes with all
$c(\b)$ and so it lies in $\A^+(M, 1 \ox \End W)$. In conclusion, the
quantity $F_{\BB}$ equals $\AA^2 - \dot\mu(R)$, $F_{\BB}$ is an
``internal'' curvature and a functional of $D$ whenever $R$ is. Now,
the Riemann tensor $R$ is a functional of $\Dslash$~\cite{ConnesGrav}.
Therefore $F_{\BB}$ is a functional of the pair $(D,\Dslash)$. We
henceforth write $F[D]$ for short.

It remains to compute $F[D]$ for the standard $K$-cycle. Recall that
there one has $\BB = \BB_0 + \BB_1,$ where $\BB_0$ holds the Higgs
term and $\BB_1$ contains the usual Yang--Mills terms. It is not hard
to see that $F[D] = \BB_0^2 + [\BB_1, \BB_0] + \BB_1^2$, as an
orthogonal direct sum of terms.

The representation of $\HH \oplus \C \oplus M_3(\C)$ on~$\H_F$
decomposes into representations on the lepton, quark, antilepton and
antiquark sectors: $\H_F = \H_F^+ \oplus \H_F^-
= \H_\ell^+ \oplus \H_q^+ \oplus \H_\ell^- \oplus \H_q^-$, each of
which in turn decomposes according to chirality:
$\H_\ell^+ = \H_{\ell R}^+ \oplus \H_{\ell L}^+$ and so on. For the
quark sector and the lepton sector with massless neutrinos, we have
respectively
\begin{eqnarray*}
\H_q^+ &=& (\C \oplus \C)_R \ox \C^N \ox \C^3_{\mathrm{col}}
\ \oplus \ \C^2_L \ox \C^N \ox \C^3_{\mathrm{col}},
\\
\H_\ell^+ &=& \H_{R\ell}^+ \oplus \H_{L\ell}^+
= \C_R \ox \C^N \,\oplus\, \C^2_L \ox \C^N.
\end{eqnarray*}
In this basis, and on applying the ``unimodularity
condition''~\cite{Cordelia}, the superconnection associated to $D$
corresponds to:
\begin{eqnarray*}
\BB_{0q} &=& \pmatrix{
0 & 0 & \bar\phi_2 \ox M_u^* & -\bar\phi_1 \ox M_u^* \cr
0 & 0 & \phi_1 \ox M^*_d & \phi_2 \ox M^*_d \cr
\phi_2 \ox M_u & \bar\phi_1 \ox M_d & 0 & 0 \cr
-\phi_1 \ox M_u & \bar\phi_2 \ox M_d & 0 & 0 \cr},
\\
\BB_{0\ell} &=& \pmatrix{
0 & \phi_1 \ox M^*_e & \phi_2 \ox M^*_e \cr
\bar\phi_1 \ox M_e & 0 & 0 \cr
\bar\phi_2 \ox M_e & 0 & 0 \cr},
\\
\BB_{1q\mu} &=& \pmatrix{
\del_\mu -\frac{4}{3}ia_\mu & 0 & 0 & 0 \cr
0 & \del_\mu + \frac{2}{3}ia_\mu & 0 & 0 \cr
0 & 0 & \del_\mu - \frac{1}{3}ia_\mu - ib_{1\mu}^1 & -ib_{2\mu}^1 \cr
0 & 0 & -ib_{1\mu}^2 & \del_\mu -\frac{1}{3}ia_\mu - ib_{2\mu}^2 \cr}
\ox 1_N \ox 1_3 \\
&&\qquad  - ic_\mu^a \,1_4 \ox 1_N \ox \frac{\la_a}{2},
\\
\BB_{1\ell\mu} &=& \pmatrix{
\del_\mu + 2ia_\mu & 0 & 0 \cr
0 & \del_\mu + i(a_\mu + b_{1\mu}^1) & -ib_{2\mu}^1 \cr
0 & -ib_{1\mu}^2 & \del_\mu + i(a_\mu + b_{2\mu}^2) \cr} \ox 1_N,
\end{eqnarray*}
where the $\la_a$ are the Gell-Mann matrices and $\phi_1$, $\phi_2$
denote the (normalized) components of the Higgs field.

The rest is routine; actually what we do is only superficially
different from what is done in~\cite{IochumKS}, and we can read off
what we need as a subset of their computations. (There are a few
misprints in that reference, but they do not affect the final
results.) Finally, we have:
$$
\tr F[D]^2 = C_H \phi^4 + C_{YMH} |D_\mu\phi||D^\mu\phi|
+ C_{YM}(F_{\mu\nu}F^{\mu\nu})_{YM},
$$
where
$$
D_\mu := \del_\mu - \tihalf g_1a_\mu - \tihalf g_2\,\tau \cdot b_\mu
$$
with an obvious notation. Our nonperturbative approach gives, for the
surviving terms, \emph{exactly the same coefficients} $C_H$,
$C_{YMH}$, $C_{YM}$ as the CC Lagrangian. We shall not bother to write
them down.

\subsection*{A particular action functional}

It should be noted that $F[D] \neq F_\nc$. The missing term in $F[D]$
is the mass term in the Higgs sector. (Actually, without fermion
families replication, the whole Higgs sector in $F_\nc$ is simply
zero. The present integral formulation eliminates this quirk of the
differential one, at the price of withdrawing the tentative claim of a
NCG-based explanation for such replication.) That missing term is
provided by the already considered $\ncint D^2\,ds^4$ term, that gives
us, besides the Einstein--Hilbert Lagrangian, the term in the square
of the Higgs field: both pieces of the puzzle fit together!

In conclusion, the bosonic action is schematically written as
\begin{equation}
\ncint (D^2 + F[D]^2) \,ds^4.
\label{CostaRicaAction}
\end{equation}
We must allow in the first summand an \emph{a priori} indeterminate
length scale $l$, for dimensional reasons; and an indeterminate
numerical coefficient $g$ in the second. Therefore:
\begin{eqnarray*}
B[D] &=& \ncint (l^{-2} D^2 + g^2 F[D]^2) \,ds^4
\\
&=& \frac{1}{8\pi^2} \Wres(l^{-2} D^2 + g^2 F[D]^2) \,D^{-4}
\\
&=& \frac{1}{8\pi^2} \Wres l^{-2} D^{-2} + g^2 \int_M (\tr F[D]^2).
\end{eqnarray*}

The action \eqn{CostaRicaAction} is quite simple and has a very
familiar look. There is a ``kinetic'' term given by the square of the
derivative (momentum) term. This term provides the action more
intrinsically connected with the nature of spacetime. Then, in the
presence of an internal structure, there is a ``potential'' term,
which is quadratic, another familiar occurrence.

As argued in~\cite{CarminatiIKS}, one can get more freedom by allowing
the quark and the lepton sectors to enter with different coefficients.
This redefinition of the noncommutative integral is permissible by the
existence of a superselection rule.

Several matters remain to be addressed. One can try to give for the
$F[D]$ term a more algorithmic expression. One can, perhaps, by
adjusting properly the theory to the known Standard Model parameters,
indulge in a new round of that favourite pastime of noncommutative
geometers, Higgs particle mass speculation. Nevertheless, comparison
of the resulting parameters with experimental values is hampered
because, as noted before, we still have no good theoretical argument
for fixing the intermediate scale at which the NCG constraints make
sense. Unless such theoretical input can be found, it rather looks as
though we shall have to wait for the experiment eventually to tell us
at what scale the present product algebra structure is likely to break
down. Last, but not least, one needs a deeper understanding of the
apparent need to apply the spectral action principle to the standard
$K$-cycle given by $D$, as opposed to the ``physical'' Dirac operator
$PDP$.

\subsubsection*{Acknowledgments}

HF, JMG-B and JCV acknowledge support from the Universidad de Costa
Rica; JMG-B also thanks the Departamento de F\'{\i}sica Te\'orica de
la Universidad de Zara\-goza, JCV the Centre de Physique Th\'eorique
(CNRS--Luminy) and FL the Department of Theoretical Physics at Oxford
for their hospitality. FL would especially like to thank all the
members of the Mathematical Physics Group at the UCR for the splendid
hospitality extended to him in Costa Rica. We thank A. Connes and A.
Rivero for discussions and E. Alvarez, M. Asorey and A. Galindo for
helpful suggestions.

\end{document}